\documentclass[prl,aps,showpacs,twocolumn,superscriptaddress,amsmath,amssymb]{revtex4}

\usepackage[pdftex]{graphics,hyperref}
\usepackage{graphicx,dcolumn,bm}

\begin{document}

\title{Generation of Symmetric Dicke States of Remote Qubits with Linear Optics}

\author{C. Thiel}
\affiliation{Institut f\"ur Optik, Information und Photonik, Max-Planck Forschungsgruppe, Universit\"at Erlangen-N\"urnberg, 91058 Erlangen, Germany}

\author{J. von Zanthier}
\affiliation{Institut f\"ur Optik, Information und Photonik, Max-Planck Forschungsgruppe, Universit\"at Erlangen-N\"urnberg, 91058 Erlangen, Germany}

\author{T. Bastin}
\affiliation{Institut de Physique Nucl\'eaire, Atomique et de Spectroscopie, Universit\'e de Li\`ege, 4000 Li\`ege, Belgium}

\author{E. Solano}
\affiliation{Physics Department, ASC, and CeNS, Ludwig-Maximilians-Universit\"at, 80333 Munich, Germany}
\affiliation{Secci\'on F\'isica, Departamento de Ciencias, Pontificia Universidad Cat\'olica del Per\'u, Lima, Peru}

\author{G. S. Agarwal}
\affiliation{Department of Physics, Oklahoma State University, Stillwater, OK 74078-3072, USA}

\date{\today}

\begin{abstract}
We propose a method for generating all symmetric Dicke states, either in the long-lived internal levels of {\em N} massive particles or in the polarization degrees of freedom of photonic qubits, using linear optical tools only. By means of a suitable multiphoton detection technique, erasing {\it Welcher-Weg} information, our proposed scheme allows the generation and measurement of an important class of entangled multiqubit states.
\end{abstract}

\pacs{03.67.Mn,32.80.Pj,03.67.-a}

\maketitle

Multipartite entanglement is arguably at the center of interest of most fields related to entanglement and quantum information theory. Unfortunately, its characterization is neither fully understood nor completed and, at the moment, we only know how to classify the entanglement of a few qubits~\cite{Duer:2000:a,Verstraete:2002:a,Lamata:2007:a}. However, these drawbacks have not prevented the apparition of a number of proposals for generating and measuring entangled states, besides their possible applications.

The efficient and scalable preparation of entangled multiqubit states is a key ingredient for the further characterization and experimental study of multipartite entanglement. Several experiments have already observed genuine entangled multiphoton states~\cite{Bouwmeester:1999:a,Kiesel:2006:a} as well as entangled distant atomic states~\cite{Julsgaard:2001:a,Matsukevich:2006:a,Chou:2005:a}. While some of the latter experiments are based on the exchange of photons between the qubits, there are other proposals for projecting distant non-interacting particles into entangled states via photonic measurements~\cite{Cabrillo:1999:a,Bose:1999:a,Skornia:2001:a,Simon:2003:a,Duan:2003:a,Duan:2001:a}. Further the very recent experiments observing interference of light emitted by two atoms~\cite{Beugnon:2006:a,Maunz:2007:a,Maunz:2007:b} make use of projective measurements. Hereby, the important class of Dicke states~\cite{Dicke:1954:a} represents a particular interesting set of quantum states associated with high robustness against particle loss~\cite{Stockton:2003:a,Bourennane:2006:a} and non-local properties of genuine entangled multipartite states~\cite{Toth:2005:a,Usha:2007:a,Retzker:2007:a}. Recently, the entangled symmetric Dicke state $|2,0\rangle$ of four photonic qubits was studied in an experiment involving linear optics only~\cite{Kiesel:2006:a}. In this experiment, among other features, the possibility of generating both classes of tripartite entangled states by projecting one of the four qubits was observed.

In this letter, we propose a method for generating any symmetric Dicke state either in distant matter or in photon polarization qubits using a multifold detection technique. In this case, we grant access to the generation and measurement of this important class of genuine entangled states for potentially any number $N$ of qubits. Our method relies on the far-field detection of $N$ photons incoherently emitted by $N$ initially excited atoms via spontaneous decay using suitably oriented polarizers. Unlike former proposals for entangling distant qubits based on projective measurements~\cite{Cabrillo:1999:a,Bose:1999:a,Simon:2003:a,Duan:2003:a}, our scheme uses explicitly the geometrical phase differences between the possible quantum paths. Furthermore, using a complementary technique, we show how to generate any symmetric Dicke state in the polarization degree of freedom of photon qubits.

In an $N$ spin-$\frac{1}{2}$ compound system, the Dicke states, usually denoted by $|S,m\rangle$, are defined as the simultaneous eigenstates of both the square of the total spin operator $\hat{\bf S}^2$ and its $z$-component $\hat{S}_z$, where $S(S+1)\hbar^2$ and $m\hbar$ are the corresponding eigenvalues~\cite{Mandel:1995:a}. The ${N+1}$ states with the highest value of the {\em cooperation number} $S=N/2$ form a special subset of all $2^N$ Dicke states. These states $|\frac{N}{2},m\rangle$ are the only ones which are totally symmetric under permutation of any particles and are usually written as
\begin{eqnarray}\label{Nstate}
\lefteqn{{\textstyle|\frac{N}{2},m\rangle}=}\nonumber\\ &\hspace{-0.35cm}\dbinom{N}{\frac{N}{2}+m}^{\!\!-\frac{1}{2}}\!\sum\limits_k{P_k(|1_1,1_2,...,1_{\frac{N}{2}+m},0_1,0_2,...,0_{\frac{N}{2}-m}}\rangle),\hspace{0.3cm}&
\end{eqnarray}
where $\{P_k\}$ denotes the complete set of all possible distinct permutations of the qubits.

Our scheme considers $N$ particles, e.g. atoms, in a $\Lambda$-configuration with upper state $|e\rangle$ and lower states $|0\rangle$ and $|1\rangle$. We may identify those states with the Zeeman-sublevels ${|e\rangle:=|e,m\!=\!0\rangle}$, ${|0\rangle:=|g_0,m\!=\!-1\rangle}$ and $|1\rangle:={|g_1,m\!=\!+1\rangle}$. The excited state $|e\rangle$ has two decay channels, $|e\rangle\rightarrow|0\rangle$ and $|e\rangle\rightarrow|1\rangle$, accompanied by the spontaneous emission of a $\sigma^+$ ($\sigma^-$)-polarized photon. For a single atom, the polarization state of the emitted photon is entangled with the corresponding ground state of the de-excited atom~\cite{Blinov:2004:a,Volz:2006:a} so that the total state of atom and photon can be written as 
\begin{eqnarray}\label{atomphoton}
|\Phi\rangle=c_0|0\rangle|\sigma^+\rangle+c_1|1\rangle|\sigma^-\rangle,
\end{eqnarray}
where $c_i$, $i = 0, 1$, is the corressponding normalized Clebsch-Gordan coefficient of the transition $|e\rangle\rightarrow|i\rangle$.

We assume the $N$ atoms to be regularly arranged in a row with equal spacing $d$ and initially excited into the upper state $|e\rangle$ by a collective laser $\pi$-pulse. $N$ detectors placed at distinct positions ${\bf r}_n$ ($n=1,\,...,\,N$) in the far-field region of the atoms, detect the spontaneously emitted photons. In front of each of the detectors, a polarization analyzer enables to measure the polarization state of the photons. Via post-selection only those events where all detectors register one and only one photon will be accepted as a measurement. For ${N=1}$, after a detector has recorded the emitted photon with a polarization equal to $\sigma^+$ ($\sigma^-$), the corresponding atom has been projected into the ground state $|0\rangle$ ($|1\rangle$). However, for ${N>1}$, the detectors located in the far-field region of the atoms are unable to distinguish which particular atom has emitted a registered photon. Therefore, after the detection of a first photon, all atoms will form a correlated state~\cite{Cabrillo:1999:a,Bose:1999:a,Simon:2003:a,Duan:2003:a}.  

The entanglement of the atoms is a consequence of two ingredients: the impossibility of the detectors to determine which atom emitted a particular photon together with the projection postulate which states that after the detection of a photon the state of the atoms is projected into a state compatible with the outcome of the measurement~\cite{Cabrillo:1999:a}. In the following we introduce a third ingredient to this scheme. It exploits the geometrical phase differences of the $N!$ quantum paths resulting from the $N!$ possibilities that each of the $N$ atoms emits a photon which is subsequently registered by one of the $N$ detectors. As will be shown below, these geometrical phase differences will allow to prepare Dicke states of arbitrary symmetric configuration.

\begin{figure}
\includegraphics[width=0.40\textwidth]{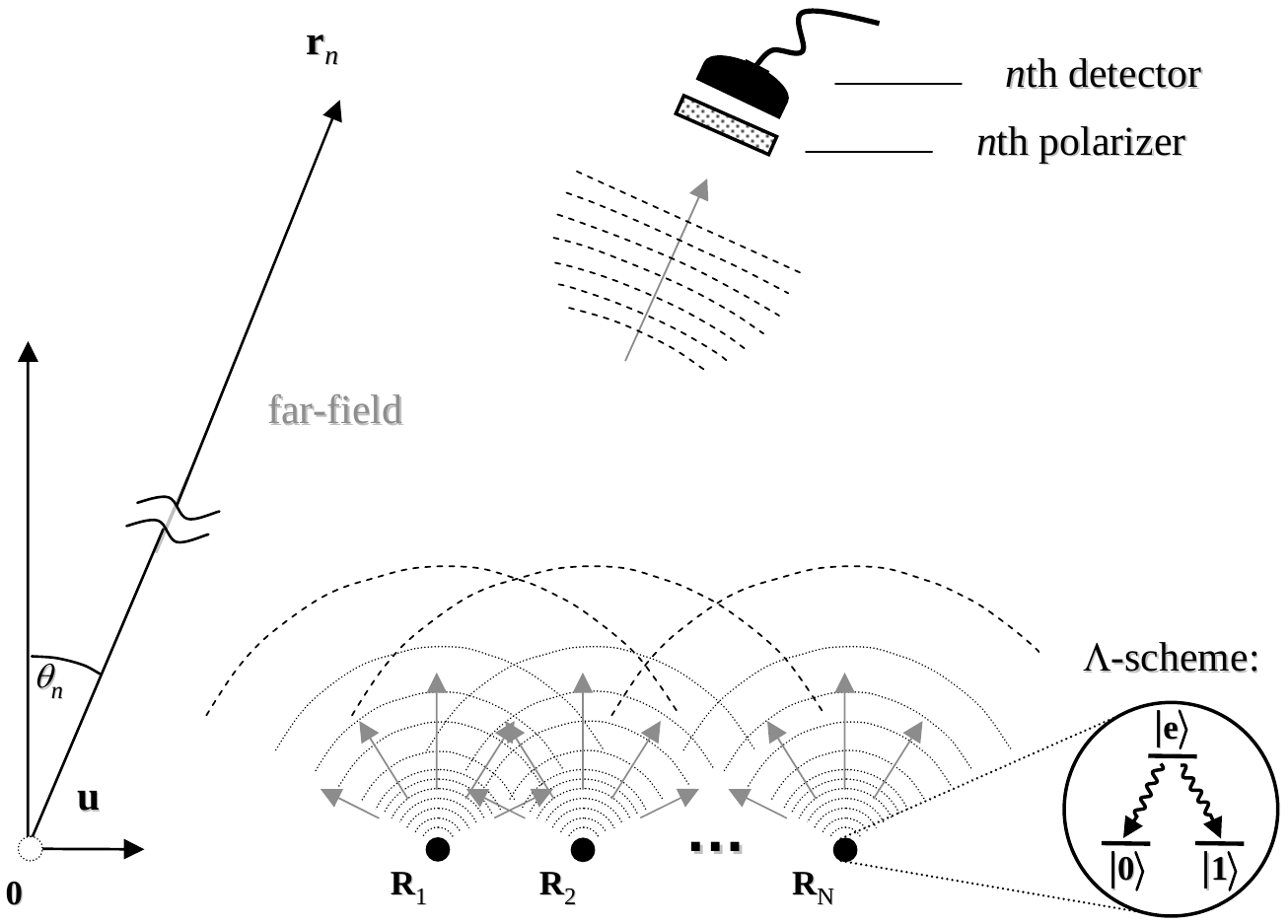}
\caption{\label{fig1} {\em N} atoms are regularly aligned in a row with spacing $d$. The origin of the coordinate system is chosen to be at one of the virtual extensions of this alignment. The $n$th detector is placed at ${\bf r}_n$ in the far-field region of the atoms, where it sees all atoms under an angle $\theta_n$ with respect to a line perpendicular on the symmetry axis of the alignment.}
\end{figure}

To show this in more detail, let us introduce the convenient coordinate system displayed in Fig.~\ref{fig1}.
As can be seen from the figure, the position of the $j$th atom (${j=1,\,...\,,N}$) is given by ${\bf R}_j=jd\,{\bf u}$, where ${\bf u}$ is a unit vector along the axis of the atoms. Denoting the unit vector along the direction of the $n$th detector by ${\bf e}_n:={\bf r}_n/r_n$, we introduce the angle $\theta_n$ shown in Fig.~\ref{fig1} so that ${\bf R}_j\cdot{\bf e}_n=j\,d\,\sin\theta_n.$ The phase difference $\delta_n$ between two photons of wavenumber $k$, emitted by adjacent atoms and both detected at ${\bf r}_n$, can then be written as
\begin{eqnarray}\label{phases}
\delta_n:=k\,({\bf R}_{j+1}\cdot{\bf e}_n-{\bf R}_j\cdot{\bf e}_n)=k\,d\,\sin\theta_n.
\end{eqnarray}

Initially, all $N$ atoms are excited into the upper state $|e\rangle$. The initial state $|\Psi_N^i\rangle$ of the atoms is thus given by
\begin{eqnarray}\label{inistate}
|\Psi_N^i\rangle=|e,e,\,...\,,e\rangle_N,
\end{eqnarray}
where the dimension of the state is indicated by the subscript $N$. The $N$ photons, subsequently emitted by the $N$ atoms, are detected by $N$ detectors at ${\bf r}_n$, $n=1,\,...,\,N$. Eventually, all $N$ atoms have thus been projected into a ground state. Hereby, each detection event has to take into account that one (unknown) atom out of $N$ possible scatterers has emitted the photon. This leads for each detection event to $N$ possible quantum paths where each of them is associated with a particular phase~\cite{Skornia:2001:a,Thiel:2006:a}. Using the coordinate system of Fig.~\ref{fig1}, the (unnormalized) operator describing the detection event of the $n$th photon at ${\bf r}_n$ can thus be written in the form~\cite{Agarwal:1974:a,Cabrillo:1999:a,Skornia:2001:a}
\begin{eqnarray}\label{detec}
\hat{D}_n:=\hat{D}_n(\delta_n,x_n)=\sum\limits_{j=1}^Ne^{ij\delta_n}|x_n\rangle_j\langle e|,
\end{eqnarray}
where $\delta_n$ is the phase introduced in Eq.~(\ref{phases}). The operator $|x_n\rangle_j\langle e|$ projects the $j$th atom from state $|e\rangle$ to the ground-state $|x_n\rangle\in\{|0\rangle,|1\rangle\}$, depending on the polarization of the photon as measured by the polarization analyzer in front of the detector.

With the detector operator of Eq.~(\ref{detec}) we can describe the detection processes of all $N$ photons emitted by the $N$ atoms. As an example, let us consider the case of $N=3$ qubits. After a first photon is detected at ${\bf r}_1$, we obtain from Eqs.~(\ref{inistate}) and (\ref{detec}):
\begin{eqnarray}\label{0ee}
\hat{D}_1\,|\Psi_3^i\rangle=e^{i\delta_1}|x_1,e,e\rangle+e^{i2\delta_1}|e,x_1,e\rangle+e^{i3\delta_1}|e,e,x_1\rangle.
\end{eqnarray}

The detection of the second and third photon may occur at ${\bf r}_2$ and ${\bf r}_3$ and we describe these events by applying successively the two detector operators $\hat{D}_2$ and $\hat{D}_3$ on the intermediate state $\hat{D}_1\,|\Psi_3^i\rangle$. The final state $|\Psi_3^f\rangle$ of the three atoms can then be written as: 
\begin{eqnarray}\label{3general}
\lefteqn{|\Psi_3^f\rangle=\hat{D}_3\,\hat{D}_2\,\hat{D}_1\,|\Psi_3^i\rangle=}\nonumber\\
&&e^{i\delta_1+i2\delta_2+i3\delta_3}|x_1,x_2,x_3\rangle+e^{i\delta_1+i2\delta_3+i3\delta_2}|x_1,x_3,x_2\rangle+\nonumber\\
&&e^{i\delta_2+i2\delta_1+i3\delta_3}|x_2,x_1,x_3\rangle+e^{i\delta_3+i2\delta_1+i3\delta_2}|x_3,x_1,x_2\rangle+\nonumber\\
&&e^{i\delta_2+i2\delta_3+i3\delta_1}|x_2,x_3,x_1\rangle+e^{i\delta_3+i2\delta_2+i3\delta_1}|x_3,x_2,x_1\rangle.\hspace{4mm}
\end{eqnarray}

For three equidistant atoms this is the most general expression of the final state. As can be seen from Eq.~(\ref{3general}), the geometrical phase differences $\delta_n$, $n=1,\,...\,,3$, determine the symmetry of the state. In particular, to generate the symmetric Dicke states $|\frac{3}{2},m\rangle$, the phases $\delta_n$ should adopt multiple values of $2\pi$, which can be determined by a suitable localization of the $N$ detectors according to Eq.~(\ref{phases}). Note that the final form of the state~(\ref{3general}) depends eventually on the orientation of the polarization analyzers in front of the  detectors: If the $n$th polarizer is oriented to transmit $\sigma^+$ ($\sigma^-$)-polarized light, the internal levels of the atoms will be projected onto the state $|x_n\rangle=|0\rangle$ ($|x_n\rangle=|1\rangle$). In particular, this means that we can generate all four symmetric Dicke states 
\begin{eqnarray}\label{3states}
 {\textstyle|\frac{3}{2},+\frac{3}{2}\rangle}\;= \hspace{1cm} & |1,1,1\rangle & \nonumber\\
 {\textstyle|\frac{3}{2},+\frac{1}{2}\rangle}\;= \hspace{1cm} & 3^{-\frac{1}{2}}\left(|1,1,0\rangle+|1,0,1\rangle+|0,1,1\rangle\right) & \nonumber\\
 {\textstyle|\frac{3}{2},-\frac{1}{2}\rangle}\;= \hspace{1cm} & 3^{-\frac{1}{2}}\left(|1,0,0\rangle+|0,1,0\rangle+|0,0,1\rangle\right) & \nonumber\\
 {\textstyle|\frac{3}{2},-\frac{3}{2}\rangle}\;= \hspace{1cm} & |0,0,0\rangle. &
\end{eqnarray}
The simple product state $|\frac{3}{2},+\frac{3}{2}\rangle$ ($|\frac{3}{2},-\frac{3}{2}\rangle$) can be obtained by orienting the three polarizers to transmit $\sigma^-$ ($\sigma^+$)-polarized light so that all atoms are projected onto the state $|1\rangle$ ($|0\rangle$). It is, however, also possible to generate the genuine tripartite entangled state $|\frac{3}{2},+\frac{1}{2}\rangle$ ($|\frac{3}{2},-\frac{1}{2}\rangle$). In this case, one polarizer should be oriented to transmit $\sigma^+$ ($\sigma^-$)-polarized and two polarizers to transmit $\sigma^-$ ($\sigma^+$)-polarized light. Hereby it does not matter which of the three detectors is actually  sensitive to $\sigma^+$- or $\sigma^-$-polarized photons, since all detectors are placed in the far-field region of the atoms and in a symmetric formation where the phases $\delta_n$ are equal to multiples of $2\pi$.

So far we showed how to generate all the symmetric Dicke states for ${N=3}$ atoms. The generalization to an arbitrary number $N$ of atoms is nevertheless straightforward. For this, we have to place again all $N$ detectors at positions ${\bf r}_1,\,...\,,{\bf r}_N$ such that the phases $\delta_n$ adopt multiple values of $2\pi$. The state of the $N$ atoms after a first photon has been detected at ${\bf r}_1$ can be calculated by applying the operator $\hat{D}_1$ on the initial state~(\ref{inistate}). From this we obtain
\begin{eqnarray}\label{iniN}
\hat{D}_1|\Psi_N^i\rangle=\sum\limits_kP_k(|x_1,e,\,...\,,e\rangle_N),
\end{eqnarray}
where $\{P_k\}$ denotes the set of all possible permutations of the $N$ qubits.

In analogy to the case ${N=3}$, we assume that the $N-1$ remaining photons are detected at positions ${\bf r}_2,{\bf r}_3,\,...\,,{\bf r}_N$, respectively. We can calculate the final state of the atoms, after all $N$ photons have been detected at the $N$ detectors, by applying the $N-1$ detector operators $\hat{D}_2,\,\hat{D}_3,\,...,\,\hat{D}_N$ on the intermediate state~(\ref{iniN}). From this we obtain:
\begin{eqnarray}\label{Nstates}
|\Psi_N^f\rangle=\sum\limits_kP_k(|x_1,x_2,\,...\,,x_N\rangle_N).
\end{eqnarray}

With the final state of the $N$ atoms given by Eq.~(\ref{Nstates}), we still have to choose the orientation of the $N$ polarizers to determine the final state of the $N$ qubits $|x_n\rangle$. For example, if we want to generate the symmetric Dicke state $|\frac{N}{2},m\rangle$, with $m \in-\frac{N}{2},...,\frac{N}{2}$, we have to choose ${\frac{N}{2}+m}$ polarizers to be sensitive to $\sigma^+$-polarized light and ${\frac{N}{2}-m}$ polarizers to be sensitive to $\sigma^-$-polarized light; this will determine the final state of the atoms to contain $\frac{N}{2}+m$ qubits in the state $|1\rangle$ and $\frac{N}{2}-m$ in the state $|0\rangle$. Again assuming that each detector registers one and only one photon, the atoms are projected into the state $|\frac{N}{2},m\rangle$ containing all symmetric Dicke states for an arbitrary number of particles. This outcome corresponds to the state expressed in Eq.~(\ref{Nstate}).

In principle, our method does not require nearby particles since we do not make use of any interaction between the atoms. Nevertheless the far-field condition inherent in expression (\ref{phases}), i.e., in Eqs.~(\ref{detec})-(\ref{Nstates}), implies a practical limit for the spacing of the particles. However, this limit can be overcome by using optical fibers. Linking each of the $N$ atoms with all $N$ detectors by using $N^2$ identical fibers leads as well to the $N!$ possible quantum paths discussed above. Hereby, the optical phases are no longer determined by the condition~(\ref{phases}) but simply by the optical paths between each ion and its light collecting fibers. Placing all fibers at the same distance to the ions, the condition $\delta_i=2\pi$ is thus fulfilled. Note that optical fibers are commonly used in experiments involving single atoms to collect the light of selective modes, see e.g.~\cite{Maunz:2007:b,Volz:2006:a}. In this way we can apply our scheme even to spatially far distant, i.e.~{\em remote}, particles.

Finally, let us estimate the expected fidelity of our scheme, e.g., for generating the symmetric Dicke state ${|2,0\rangle}$ using ${N=4}$ adjacent atoms. In the case of ions localized in a linear trap, we assume the atoms 5~$\mu$m apart and confined to 5~nm in the lateral direction, i.e.\ perpendicular to the trap axis. Furthermore, we allow for an azimuthal detection window of $0.6^\circ$. All these uncertainties were included in our analysis via error propagation, and we estimate a fidelity of about $90\%$ for the generation of the four qubits state $|2,0\rangle$. Remarkably it was shown recently that a fidelity of $66\%$ is already sufficient to demonstrate the entanglement of this state~\cite{Toth:2005:a}. In an experiment that uses CCD-cameras covering a fair area in the detection plane, and taking into account all sources of errors mentioned above, we moreover expect the counting rate of the needed four-fold coincident events to be a few tenths of Hz with an excitation rate of several tens of MHz~\cite{Blinov:2004:a} (see~\cite{Thiel:2007:a}). In general the counting rate decreases with the number of qubits. This might limit the scalability of our scheme as is indeed the case with other experiments observing entangled photons~\cite{Bouwmeester:1999:a,Kiesel:2006:a} as well as entangled atoms~\cite{Blinov:2004:a,Volz:2006:a,Maunz:2007:b}.

In the last part of this paper we want to discuss how our method can also be used to prepare symmetric Dicke states in the polarization degree of freedom of photon qubits. Recently the Dicke state $|2,0\rangle$ has been observed as an entangled photon polarization state in a post-selective manner, by using initially entangled photons generated in SPDC~\cite{Kiesel:2006:a}. To prepare arbitrary symmetric Dicke states of photon polarization qubits we have to place the polarization analyzers, formerly positioned in front of the detectors (see Fig.~\ref{fig1}), in front of the atoms such that the polarization of each spontaneously emitted photon is determined by an individual polarizer. The setup remains otherwise identical to the one presented above: All $N$ atoms are initally prepared in the excited state $|e\rangle$ and, via post-selection, we assure that one and only one photon is registered at each of the $N$ detectors. However, after the detection of the photons the internal state of each atom is now uniquely determined by the orientation of the polarizer, i.e., in correspondence to the polarization state of the photon emitted by this particular atom. Since the photons are still detected in the far-field region of the atoms, we do not acquire Welcher-Weg information of individual photons and thus cannot determine the polarization state of each individual photon at any of the $N$ detectors. Instead, all $N$ quantum paths associated with the $N$ possibilities that a photon has been emitted by one of the $N$ atoms will contribute to a single photon detection event at a particular detector.

Introducing the wave vectors of the $N$ different spatial modes ${\bf k}_n=k\,{\bf e}_n$, defined by the unit vectors ${{\bf e}_1,\,...\,,{\bf e}_N}$ of the positions of the $N$ detectors, we only know after the detection of all $N$ photons at ${{\bf r}_1,\,...\,,{\bf r}_N}$ that each single mode ${\bf k}_n$ was populated by exactly one photon. But what was the polarization state of the photon in the $n$th mode? We define the polarization state of a photon in the mode ${\bf k}_n$ as $|x_n\rangle=|\sigma^+\rangle$ ($|\sigma^-\rangle$). Using the same detector positions as for generating Dicke states of massive particles, we obtain the same state as given in Eq.~(\ref{Nstates}), however now for the polarization state of the $N$ photons in the $N$ spatial modes. It is thus possible to generate an arbitrary symmetric Dicke state $|\frac{N}{2},m\rangle$ of photon polarization qubits, by choosing $\frac{N}{2}+m$ polarizers to be sensitive to $\sigma^-$- and $\frac{N}{2}-m$ polarizers to $\sigma^+$-polarized light.

In conclusion, we have demonstrated that it is possible to generate all symmetric Dicke states for distant matter as well as for photon polarization qubits using linear optical tools only. Our method offers a simple access to genuine entangled states of any number of qubits exploiting absence of Welcher-Weg information and polarization sensitive far-field detection of photons spontaneously emitted by atoms in a $\Lambda$-configuration. As for the technical feasibility of making use of optical phase differences between single ions, we refer to~\cite{Eichmann:1993:a} where first order interferences of light coherently scattered by two ions were observed. It can be seen from Eq.~(\ref{3general}) that our method is also capable of generating entangled quantum states different from the symmetric Dicke states. In addition, considering more general atomic arrangements or orientations of the polarizers it is possible to generate other classes of entangled states~\cite{Thiel:2007:a}.

We gratefully acknowledge financial support by the Dr. Hertha und Helmut Schmauser foundation. G.S.A. thanks NSF grant no NSF-CCF-0524673 for supporting this collaboration. E.S. thanks finnancial support of EuroSQIP and DFG SFB 631 projects.

\end{document}